\documentclass{PoS}
\usepackage{cite}

\title{
{\small{DESY 13--149, DO-TH 13/21, SFB/CPP-13-58, LPN 13-057}}
\\
ABM news and benchmarks}

\ShortTitle{ABM news}

\author{\speaker{Sergey Alekhin}\thanks
{This work has been supported in part 
by Helmholtz Gemeinschaft under contract VH-HA-101 ({\it Alliance Physics at the Terascale}), 
DFG Sonderforschungsbereich/Transregio~9 and by the European Commission through contract      
PITN-GA-2010-264564 ({\it LHCPhenoNet})}\\
        DESY, Platanenallee 6, D--15738 Zeuthen, Germany;\\
Institute for High Energy Physics,142281 Protvino, Moscow region, Russia\\
        E-mail: \email{sergey.alekhin@desy.de}}

\author{Johannes Bl\"umlein\\
        DESY, Platanenallee 6, D--15738 Zeuthen, Germany\\
        E-mail: \email{johannes.bluemlein@desy.de}}

\author{Sven-Olaf Moch\\
II. Institut f\"ur Theoretische Physik, Universit\"at Hamburg
    Luruper Chaussee 149, D-22761 Hamburg, Germany;\\
        DESY, Platanenallee 6, D--15738 Zeuthen, Germany\\
        E-mail: \email{sven-olaf.moch@desy.de}}

\abstract{We report on progress in the determination of the unpolarised 
nucleon PDFs within the ABM global fit framework. The data used in the 
ABM analysis are updated including the charm-production and the high-$Q^2$ 
neutral-current samples obtained at the HERA collider, as well as the LHC data 
on the differential Drell-Yan cross-sections. An updated set of the PDFs with 
improved experimental and theoretical accuracy at small $x$ is presented. 
We find minimal impact of the $t$-quark production cross section measured
at the Tevatron and the LHC on the gluon distribution and the value of the 
strong coupling constant $\alpha_s$ determined from the ABM fit in the case
of the $t$-quark running-mass definition. In particular, the value of 
$\alpha_s(M_Z)=0.1133\pm 0.0008$ is obtained from the variant of the 
ABM12 fit with the Tevatron and CMS $t$-quark production cross-section data 
included and the $\overline{\rm MS}$ value of $m_t(m_t)=162~{\rm GeV}$.   
}

\FullConference{XXI International Workshop on Deep-Inelastic Scattering and Related Subjects - DIS2013,\\
		22-26 April 2013\\
		Marseilles,France}

\begin{document}

Recent progress in the analysis of the collider data allows a gradual improvement of the 
PDF accuracy at small $x$ being of particular importance for the phenomenology at the LHC. Since 
the release of the ABM parton distribution functions (PDFs)~\cite{Alekhin:2012ig}  a 
new combined HERA data set on semi-inclusive charm production in deep-inelastic scattering (DIS)  
has been obtained~\cite{Abramowicz:1900rp}. It provides a complementary constraint on the 
gluon distribution and allows to benchmark the factorization schemes employed for the description 
of the heavy-quark contribution to DIS. In addition, the first LHC data on the differential 
distributions of the charged leptons produced in the Drell-Yan (DY) 
process~\cite{Aad:2011dm,Chatrchyan:2012xt,Aaij:2012vn,lhcbe} allow to check the PDFs tuned to 
the fixed-target data at values of the Bjorken variable $x \sim 0.01$ and factorization scales 
$\mu\sim 100 ~{\rm GeV}$. In these proceedings we discuss an update of the ABM11 analysis including 
these HERA and LHC data sets. We also add to the analysis 
the HERA neutral-current 
data with $Q^2>1000~{\rm GeV^2}$ omitted earlier in the ABM11 fit. The
theoretical footing is correspondingly developed by 
accounting for the contribution due to  
the $Z$-boson exchange. We also update the next-to-next-to-leading-order
(NNLO) Wilson coefficients for the heavy-quark electroproduction employing
those having been derived recently by a combination of the partial NNLO results stemming 
from threshold resummation and the high-energy limit with the 
Mellin moments \cite{Bierenbaum:2009mv}
of the massive operator-matrix elements, cf.~\cite{Kawamura:2012cr}. 
The $t$-quark production at the LHC~\cite{lhct} and Tevatron~\cite{tevt}
\begin{table}[th!]
\renewcommand{\arraystretch}{1.3}
\begin{center}                   
{\small                          
\begin{tabular}{|c|c|c|c|c|}   
\hline                           
{Experiment}                      
&ATLAS~\cite{Aad:2011dm}                         
&{CMS~\cite{Chatrchyan:2012xt}}  
&{LHCb~\cite{Aaij:2012vn}}
&{LHCb~\cite{lhcbe}}                            
\\                                                        
\hline
{Final states}                                                   
& $W^+\rightarrow l^+\nu$
& $W^+\rightarrow e^+\nu$
&$W^+\rightarrow \mu^+\nu$
& $Z\rightarrow e^+e^-$                                                        
\\                                                        
& $W^-\rightarrow l^-\nu$
&$W^-\rightarrow e^-\nu$
&$W^-\rightarrow \mu^-\nu$
&                                                         
\\                                                        
& $Z\rightarrow l^+l^-$
&                                                         
&                                                         
&                                                         
\\                                                        \hline                                                    
{Luminosity (1/pb)}                      
&35                         
&840  
&{37}
&{940}                            
\\                                                        
\hline                                                    
$NDP$
&30                      
&11  
&10
&9                            
\\                                                        
\hline
 $\chi^2$
 &$34.5(7.7)$
 &$11.8(4.7)$
 &13.0(4.5)
 &11.5(4.2)
\\
\hline                                          
\end{tabular}
}
\caption{\small The value of $\chi^2$ obtained for different samples of 
the Drell-Yan LHC data with the NNLO ABM11 PDFs.
The figures in parenthesis give one standard deviation of 
$\chi^2$ equal to $\sqrt{2NDP}$.}
\end{center}
\label{tab:chi2}
\end{table}
\begin{figure}[h]
  \centering
  \includegraphics[width=0.8\textwidth,height=3.3in]{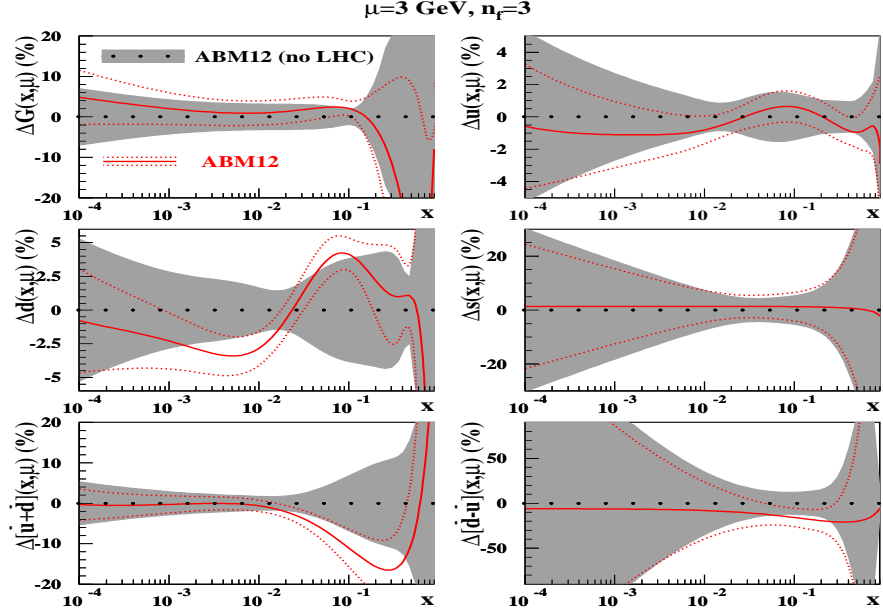}
\caption{\protect\label{fig:abm12}                                              
        \small The relative change in the 3-flavor 
ABM PDFs at the factorization scale of $\mu=3~{\rm GeV}$ due to the LHC DY 
data~\cite{Aad:2011dm,Chatrchyan:2012xt,Aaij:2012vn,lhcbe} 
(solid curves) in comparison with the uncertainties in the variant of ABM12 
fit performed without employing those data (shaded area). The uncertainties in
the ABM12 fit are displayed by the dotted curves.  
} 
\end{figure}
can potentially also constrain the PDFs, particularly the 
gluon distribution. However, this constraint is quite sensitive to the 
$t$-quark mass $m_t$. Meanwhile the experimental determination of $m_t$ is
performed on the basis of Monte-Carlo studies yet missing the high-order
corrections and its result cannot be directly used in comparisons with theoretical 
precision calculations. Furthermore, the NNLO corrections to the $t$-quark production 
cross section~\cite{Czakon:2013goa} depend on the mass definition~\cite{Dowling:2013baa}.
Therefore we check the $t$-quark data~\cite{lhct,tevt} both for the case of
the pole- and $\overline{\rm MS}$--masses at different values of $m_t$.
In the remaining part of the proceedings we discuss comparisons of the LHC 
DY--data
with the ABM11 predictions and the incorporation of those data into the ABM fit,
outline the ABM12 PDF features, and 
discuss the impact of the $t$-quark data on the ABM PDFs and the strong coupling 
constant~\footnote{The impact of the charm-production data~\cite{Abramowicz:1900rp}
and related theoretical improvements   
are described elsewhere~\cite{dis2}.}.
%
\begin{figure}[th!]
\centerline{
  \includegraphics[width=9.5cm,height=1.9in]{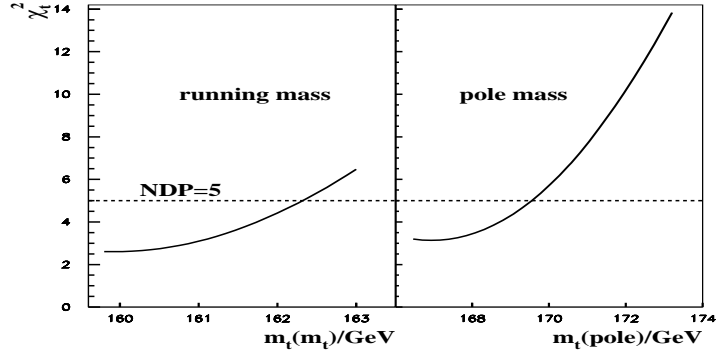}}
  \caption{\small
    \label{fig:chi2t}
      The $\chi^2$ profile for the Tevatron and LHC 
      $t\bar{t}$ cross section data~\protect\cite{lhct,tevt}
      versus the $t$-quark mass obtained in the variants 
      of ABM12 fit with those data included
      and different $t$-quark mass definitions (running mass: left, 
      pole mass: right). 
     The $NDP=5$ for this subset is displayed by the dashed line. 
}
\end{figure}

The $W$- and $Z$-boson production at the LHC has been studied by their 
leptonic decays and the most accurate data are obtained for the electron 
and muon channels in the form of differential distributions of the 
final-state charged leptons. Confronting these data with the theoretical 
predictions requires fully exclusive calculations which are implemented 
in two existing codes, DYNNLO~1.3~\cite{Catani:2009sm} and 
FEWZ~3.1~\cite{Li:2012wn}. Taking advantage of both we compute 
the prediction of the central values with DYNNLO and the PDF uncertainties with 
FEWZ. The NNLO ABM11 predictions obtained in this way are in  good agreement 
with the DY data by the ATLAS~\cite{Aad:2011dm}, 
CMS~\cite{Chatrchyan:2012xt}, and LHCb~\cite{Aaij:2012vn,lhcbe} experiments 
with account of the PDF uncertainties~\footnote{The benchmark of PDFs using 
the DY LHC data~\cite{Ball:2012wy}
is based on the NLO calculations combined with the NNLO K-factors 
and performed without taking into account the PDF uncertainties in the
statistical analysis.}.   
The values of $\chi^2$ are in a good agreement with the 
number of data points (NDP) for each LHC data sample, cf. Table~1,
and the total value of 
$\chi^2/NDP=71/60$ is comparable with 1 within its statistical fluctuations 
of $\sqrt{2/NDP}$. 
Incorporating the DY LHC data into the NNLO PDF fit in a straightforward way
requires an enormous computational power. Therefore it is commonly performed 
using the grids calculated in advance for a wide set of PDFs covering 
their expected variations. For this purpose we use the DY cross section values
calculated for 27 PDF sets encoding the ABM11 uncertainties due to
the fitted PDF parameters. In the fit, including the DY LHC data, the 
cross section value corresponding to a current value of the PDF parameters is 
computed by linear interpolation between grid values. This approach is well 
justified if the parameter variations are within their error margins. 
This holds in our case since the data are in agreement with the previous ABM11 
predictions. The change in the PDFs due to the inclusion of the LHC data in general 
is also obtained within the PDF uncertainties, cf. Fig.~\ref{fig:abm12}. The biggest 
changes are observed at $x\sim 0.1$, the region most sensitive to $W/Z$--production 
at the LHC. There the $d$-quark distribution grows by some 3\%.    
It is also worth noting that its error is reduced dramatically due to the LHC data being 
free from the impact of nuclear corrections. 
The non-strange sea quark distribution goes down even stronger, although remaining 
within the uncertainties and the change in the strange sea is marginal. 
The value of $\alpha_s(M_Z)=0.1132\pm 0.0011$ obtained in the ABM12 fit is in
a good agreement with the ABM11 value of $\alpha_s(M_Z)=0.1134\pm 0.0011$.
\begin{figure}[h]
  \centering
  \includegraphics[width=0.8\textwidth,height=2.2in]{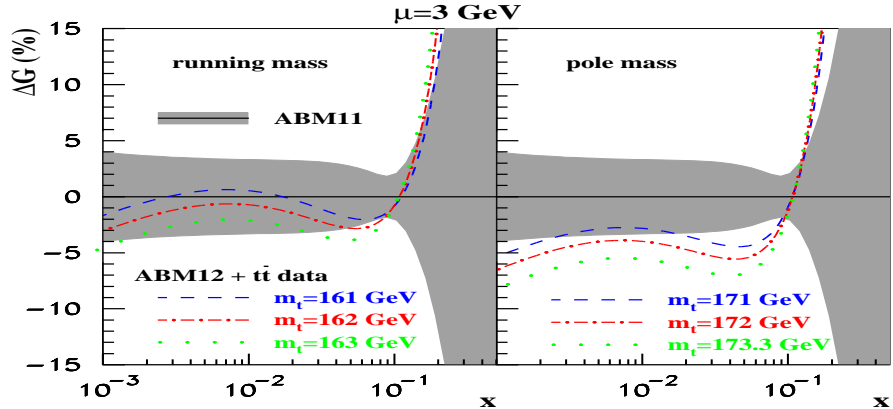}
\caption{\protect\label{fig:pdftq}                                              
        \small The relative uncertainty of the ABM11 gluon distribution in the 3-flavor
     scheme at the factorization scale of $\mu=3~{\rm GeV}$ (grey area) in comparison 
     to its relative change due to inclusion of the $t\bar{t}$ cross section 
     data with the different mass definitions: 
     running mass (left), pole mass (right),  
     and the $t$-quark mass settings as indicated in the plot.
} 
\end{figure}

The ATLAS and CMS experiments collected $t$-quark 
samples at the c.m.s. collision energy of 7 and 8 TeV and provided 
estimates of the $t \bar t$ production cross 
sections~\cite{lhct}. We
have checked the combination of these data with those of Tevatron~\cite{tevt}
in the ABM12 fit using the pole- and 
$\overline{\rm MS}$-masses for $m_t$. In both cases the 
QCD corrections up to NNLO are taken into account~\cite{Czakon:2013goa}. 
However, the running-mass definition has the advantage to provide a better 
perturbative stability~\cite{Dowling:2013baa}.
The $t$-quark data can be easily accommodated into the ABM fit, 
taking the running-mass definition, cf. Fig.~\ref{fig:chi2t}. 
In case of the pole mass the agreement is worse, particularly at the experimentally 
measured value of $m_t=173.3~{\rm GeV}$~\footnote{Note that
the benchmarking of the ABM11 PDFs with the $t$-quark data~\cite{Czakon:2013tha} 
is performed with the pole-mass definition.}. 
The impact of the $t$-quark data on the gluon distribution also depends
on the value of $m_t$ and on the mass definition, 
cf. Fig~\ref{fig:pdftq}. 
For the running-mass case it does not exceed $1\sigma$ in general, while it 
is bigger for the pole mass.
The main contribution to the $\chi^2$ value comes from
the ATLAS data, with somewhat overshoot of the ABM predictions. Furthermore, in 
the variant of the ABM12 fit including only the CMS and Tevatron $t$-quark data 
the PDFs are changed to a much smaller extend than with the ATLAS data being
included. This displays a certain tension between the ATLAS and CMS 
data and prevents from including the LHC $t$-quark data into the fit. Moreover,
essential experimental details about systematic error correlations for these 
data sets and the LHC beam energy uncertainty are still missing. 
Meanwhile, in the variant of our 
analysis including the CMS and Tevatron $t$-quark data only
and with the $\overline{\rm MS}$ value of $m_t(m_t)=162~{\rm GeV}$ 
we obtain the value of $\alpha_s(M_Z)=0.1133\pm 0.0008$. It is in very good
agreement with the result in the ABM11 fit and smaller than the value 
of $\alpha_s(M_Z)=0.1187\pm 0.0027$ obtained by the CMS collaboration 
referring to the ABM11 PDFs and using the pole-mass definition 
$m_t=173.2$~\cite{Chatrchyan:2013haa}.

\end{document}